\documentclass[conference]{IEEEtran}
\usepackage{cite}
\usepackage{amsmath,amssymb,amsfonts}
\usepackage{algorithmic}
\usepackage{graphicx}
\usepackage{textcomp}
\usepackage{xcolor}
\usepackage{fancyhdr}
\usepackage[hyphens]{url}

\usepackage{mathptmx} 

\usepackage{tikz}

\usepackage[normalem]{ulem}
\usepackage{microtype}

\usepackage{hyperref}
\usepackage{xspace}
\usepackage{amsmath,amssymb,amsfonts}
\usepackage[format=plain,font=small]{caption}
\usepackage{algorithmic}
\usepackage{graphicx}
\usepackage{graphics}
\usepackage{fancyhdr}
\usepackage{textcomp}
\usepackage{verbatim}
\usepackage{listings}
\usepackage{enumitem}
\usepackage{hhline}
\usepackage{threeparttable}
\usepackage{colortbl}
\usepackage{subcaption}
\usepackage{tabularx}
\usepackage{multirow,array}
\usepackage{booktabs}
\usepackage{soul}
\newcommand{\ignore}[1]{}
\newcommand{\system}{\textit{Cheetah}\xspace}

\newcommand{\RCG}[1]{\rowcolor[HTML]{656565}}
\newcommand{\RCC}[1]{\rowcolor[gray]{0.9}}
\definecolor{darkgrey}{HTML}{434343}

\newcommand{\WT}[1]{\color{white}}

\lstdefinestyle{mystyle}{
    commentstyle=\color{codegreen},
    keywordstyle=\color{magenta},
    emph=[1]{LOAD,GEMM,STORE},emphstyle=[1]{\color{blue}\bfseries},
    emph=[2]{LOAD_E,STORE_E,GEMM_C,ZEROIZE},emphstyle=[2]{\color{magenta}\bfseries},
    numberstyle=\tiny\color{codegray},
    stringstyle=\color{codepurple},
    basicstyle=\ttfamily\tiny,
    breakatwhitespace=false,         
    breaklines=true,                 
    captionpos=b,                    
    keepspaces=true,                 
    numbersep=3pt,
    frame=none,
    keepspaces=true,
    showspaces=false,                
    showstringspaces=false,
    showtabs=false,                  
    tabsize=2
}
\usepackage{caption}
\usepackage{enumitem}

\def\BibTeX{{\rm B\kern-.05em{\sc i\kern-.025em b}\kern-.08em
    T\kern-.1667em\lower.7ex\hbox{E}\kern-.125emX}}

\newcommand{\iscasubmissionnumber}{21}

\fancypagestyle{firstpage}{
  \fancyhf{}

  \fancyhead[C]{\normalsize{SPSL 2021 Submission
      \textbf{\#\iscasubmissionnumber} \\ Confidential Draft: DO NOT DISTRIBUTE}} 
  \fancyfoot[C]{\thepage}
}

\newcommand{\TODO}[1]{{\color{red}[TODO: #1]}}

\title{Bandwidth Utilization Side-Channel on ML Inference Accelerators}

\makeatletter
\newcommand{\linebreakand}{%
  \end{@IEEEauthorhalign}
  \hfill\mbox{}\par
  \mbox{}\hfill\begin{@IEEEauthorhalign}
}
\makeatother

\author{
\IEEEauthorblockN{Sarbartha Banerjee}
\IEEEauthorblockA{The University of Texas at Austin\\
sarbartha@utexas.edu}

\and
\IEEEauthorblockN{Shijia Wei}
\IEEEauthorblockA{The University of Texas at Austin\\
shijiawei@utexas.edu}

\and
\IEEEauthorblockN{Prakash Ramrakhyani\\}
\IEEEauthorblockA{ARM Research\\
prakash.ramrakhyani@arm.com}

\linebreakand

\IEEEauthorblockN{Mohit Tiwari\\}
\IEEEauthorblockA{The University of Texas at Austin\\
tiwari@austin.utexas.edu}
}

\begin{document}
\maketitle
\begin{abstract}
Accelerators used for machine learning (ML) inference provide great performance benefits over CPUs.
Securing confidential model in inference against off-chip side-channel attacks is critical in harnessing the performance advantage in practice. 
Data and memory address encryption has been recently proposed to defend against off-chip attacks.
    
In this paper, we demonstrate that
bandwidth utilization on the interface between
accelerators and the weight storage
can serve a side-channel
for leaking confidential ML model architecture.
This side channel is independent of the type of interface,
leaks even in the presence of data and memory address encryption
and can be monitored through performance counters
or through bus contention from an on-chip unprivileged process.\footnote{\textbf{Presented at Secure and Private Systems for Machine Learning (SPSL 2021)}}
    
\end{abstract}

\section{Introduction}


Deep learning model inference services have spawned a domain-specific computing revolution. High performance ML inference accelerators in the form of neural processing units (NPU) are being developed by both industry~\cite{tpu,ArmEthos-N,Brainwave} and academia~\cite{eie,VTA,dadiannao}.
NPUs may be integrated either in the system-on-chip (SoC)~\cite{ArmEthos-N} or connected to system bus~\cite{tpu}. Inference-as-a-service (InFaaS)~\cite{infaas} is deployed by cloud providers like Amazon Sagemaker~\cite{sagemaker} running user inference on ML accelerators.
This incentivizes ML model vendors to host trained models on cloud platforms and provide services on confidential user data like face recognition and organizational data like disease classification on patient private data. 

From the security perspective, the model vendor requires the cloud provider to protect the confidentiality of model parameters as well as the layer dimensions.
Prior work~\cite{bb1,bb2,bb3} show how knowledge of layer dimensions can be used to steal a victim's ML IP by reconstructing a model with similar accuracy.
An attacker can further use a stolen model to launch 
adversarial attacks on the victim system~\cite{adv1,adv2}.

Temporal and spatial sharing of NPUs by multiple applications
has been proposed to improve system scalability and overall inference time~\cite{prema,planaria}.
This allows multiple tenants sharing the memory bus to infer victim model utilization through bandwidth contention channels.
To support such sharing, the cloud hypervisors collect NPU resource
utilization information (e.g.~memory bus utilization) for each tenant
for providing quality-of-service (QoS) guarantees, memory traffic and infrastructure management.

In this paper, we develop a new attack based on observing the NPUs bandwidth utilization.
We highlight that observing this side channel alone can leak the ML model structure even when off-chip data and addresses are encrypted. 
Summarizing our key contributions:
\begin{itemize}
    \item We introduce the bandwidth utilization side-channel on NPUs to leak ML model dimensions.
    \item A proof-of-concept exploit on the DRAM interface against 6 image classification models and
explore several classifier to identify model layer boundaries and layer dimension.
    \item We propose possible defences to
create a demand agnostic bandwidth utilization. Software defence leads to a $1.6x$ increase in overhead while hardware countermeasures to generate constant memory traffic has $14.6\%$ to $19.3\%$ overhead.

\end{itemize}

\section{Overview and background}



\begin{figure}[t]
    \centering
    \includegraphics[width=\linewidth]{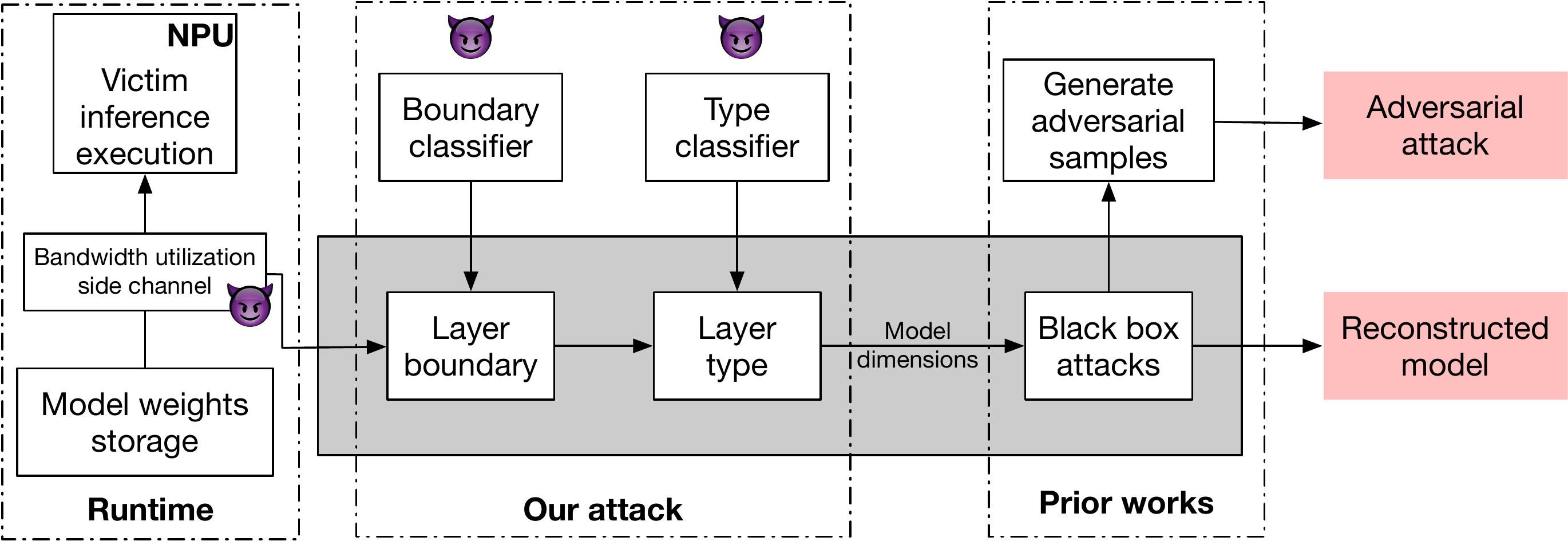}
    \caption{\textbf{Different steps of an end-to-end bandwidth utilization attack. (1) Collect bandwidth utilization; (2) Classify model layer boundary and layer type; (3) Use black box attack to reconstruct model or perform adversarial attack.}}
    \label{fig:attack_pipeline}
    \vspace{-1em}
\end{figure}

Figure~\ref{fig:attack_pipeline} shows the different stages of an end-to-end attack with utilization bandwidth as a side-channel.
The memory bus connecting the NPU and the model weight storage reveals layer variations through the bandwidth utilization side channel.
Our attack receives the bandwidth utilization trace
and performs two classification steps:
\textbf{(1)} Detect layer boundaries with a \textit{boundary detector} (Section~\ref{sec:boudarydetector});
\textbf{(2)} Split the time-series along layer boundaries to detect the layer dimensions with \textit{type classifier} (Section~\ref{sec:typeclassifier}).
The number of layers and its dimension can be fed to black box attacks for model weights reconstruction or an adversarial attack (Section~\ref{sec:black_box}).

\noindent \textbf{NPU to memory interface.}
\label{background:mem_intf}
The NPU accelerator can be resident on the system-on-chip (SoC) with memory bus connected directly to CPU last-level cache (LLC) as in ARM Ethos~\cite{ArmEthos-N} or connected with an on-chip eDRAM as in DaDianNao~\cite{dadiannao}. Other accelerators~\cite{tpu} are connected off-chip performing direct system memory accesses. 
All the model weights cannot be loaded upfront due to limited NPU internal memory.
This motivates tiling the weights for each DNN layer to fit in the NPU and 
loading them as required, therefore creating bandwidth variation
irrespective of the NPU memory interface.

\noindent \textbf{Tile size variations.}
The model layer dimensions depend on the model configuration like number of filters, height, width and channels.
The tiling is not only devised on these four dimensions but also on the data locality of the input feature maps and the intermediate partial sums. Moreover, the number and type of operation for each layer are different. Convolution and dense layers perform matrix multiplication while pooling and activation perform ALU operations like max, min or addition leading to parameter variations for each layer.

\noindent \textbf{Black box attacks.}
\label{sec:black_box}
A large body of work~\cite{bb1,bb2,bb3} recreates a model having similar classification accuracy as the victim model with only the model dimension.  As a first step, the attacker initializes a model with victim model dimension. Then, she sends multiple inference requests to the victim model and uses the classification to label the data creating a dataset. This dataset is used to train the attacker's model until it achieves similar accuracy as the victim model. prior work~\cite{adv1,adv2} further illustrates use of reconstructed model to generate adversarial examples for the victim model.

  \section{Bandwidth Utilization Attack}

\subsection{Threat Model}
We adopt a threat model similar to trusted execution environments (TEE) like Intel SGX~\cite{SGX}.
Specifically, we assume that the underlying privileged software is untrusted,
malicious co-tenant may share the same NPU
and the system may be exposed to physical attacks.
We assume the techniques in secure processors are adopted by
the NPU providers.
Beyond memory encryption and integrity checks,
we also assume memory address encryption is employed as in~\cite{Invismem,obfusmem},
thus restricting the adversary to only observe the bandwidth
utilization.

Additionally, we assume that the victim uses the optimal tile size for performance, and that the adversary can perform offline profiling to collect all possible layer and tile configuration characteristics for training her classifier(s).

\subsection{Point of Leakage}
Bandwidth utilization can be observed by malicious hypervisors
or co-executing tenants.
Cloud hypervisors collect traffic statistics for each application
for load balancing and congestion control.
Considering the threat model of a malicious cloud hypervisor,
high-precision counters can be used to
collect bandwidth utilization of each tenant.
Such widgets are (or can be) placed at
(1) the Last-Level Cache (LLC) interface,
(2) the DMA engine of the NPU,
or (3) the DRAM controller.

Unprivileged co-executing tenants share the same memory interface.
Malicious tenants in the same NPU can constantly monitor the bandwidth
and record victim utilization from bandwidth drops due to contention.
This drop in bandwidth is proportional
to the victim load request size scheduled by a physically shared NPU DMA controller.

\subsection{Example of Bandwidth Variation at DRAM Interface}
\label{sec:example_variation}
\begin{figure}[t]
    \vspace{-1em}
    \centering
    \begin{subfigure}{0.95\linewidth}
        \centerline{\includegraphics[width=\linewidth]{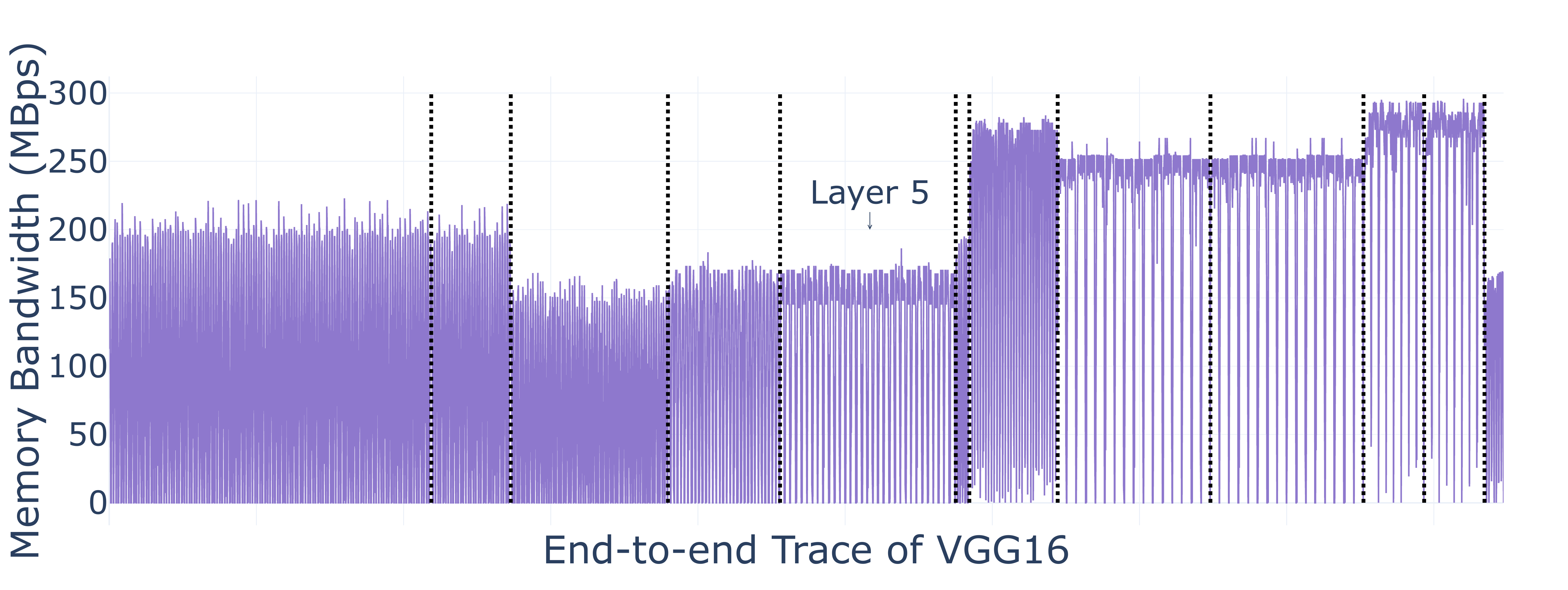}}
        \subcaption{\label{fig:vgg16_end2end}\textbf{Memory bandwidth of 12 layers in VGG16.}}
    \end{subfigure}
    \begin{subfigure}{0.95\linewidth}
        \centerline{\includegraphics[width=\linewidth]{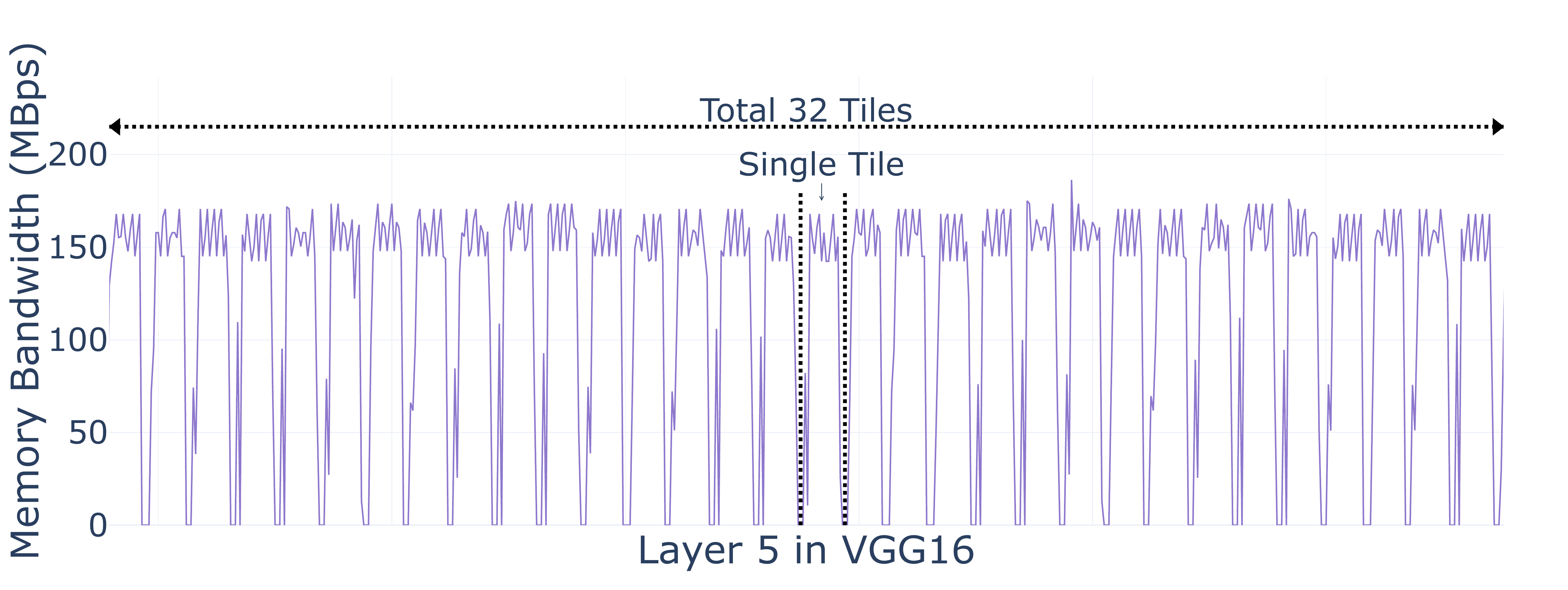}}
        \subcaption{\label{fig:vgg16_layer5}\textbf{Memory bandwidth of the 5th layer in VGG16.}}
    \end{subfigure}
    \caption{\label{fig:vgg16_attack_demo}\textbf{Different layers in vgg16 network utilizes different memory bandwidth, while bursts within each layer show number of tiles.}
    }
\end{figure}

As mentioned in Section~\ref{background:mem_intf},
tile execution of different layers on the NPU 
loads tiled matrix weights from either LLC or main memory.
Figure~\ref{fig:vgg16_end2end} presents the layer-dependent
load bandwidth variation for the model weights in a VGG16 model.
There are three types of bandwidth variations across adjacent layer boundaries:
\textbf{(T1)} Layers with different tile sizes (e.g.~layer 2 and 3) where the filter size is different with a pooling layer in between.
\textbf{(T2)} Layers with the same sized tiles but having different number of tiles (e.g.~layer 1 and 2) which is possible with adjacent layer having the same filter size but differing in the number of channels.
\textbf{(T3)} Identical adjacent layers (e.g.~layer 10 and 11) with identical sized filters having the same number of channels as in some deeper resnet layers. 

Figure~\ref{fig:vgg16_layer5} zooms into the $5^{th}$ layer
revealing the number of tiles.
Due to tile size optimization,
the number of tiles and the bandwidth utilization
vary across model layers depending on their dimensions.
There are finite tiling schemes with each having a unique signature of bandwidth and execution time. 

\subsection{Attack Demonstration}
As a proof-of-concept, we demonstrate the bandwidth utilization attack
on the DRAM interface and model the adversary as a malicious hypervisor.

\subsubsection{\textbf{Experimental Setup}}
\label{sec:expsetup}
We prototype the VTA~\cite{VTA} ML inference accelerator on a Xilinx Pynq-Z1 fpga running at 100 MHz.
VTA is an fpga implementation of an NPU with the widely used TVM~\cite{TVM} software stack.
Similar to an NPU, the VTA has its own instruction set consisting of load, compute and store operations.
We augment the VTA DMA controller with a memory transaction counter to log the read transactions. 
The model weights are not updated at inference time, hence monitoring only read transaction is enough for inferring the model dimensions. 
The counter accumulates the number of bytes transferred for each DMA transaction and stores in a memory-mapped register,
read by the runtime driver at 250KHz.

\begin{figure*}[t]
    \vspace{-1em}
    \centering
    \includegraphics[height=4.5cm,width=\linewidth]{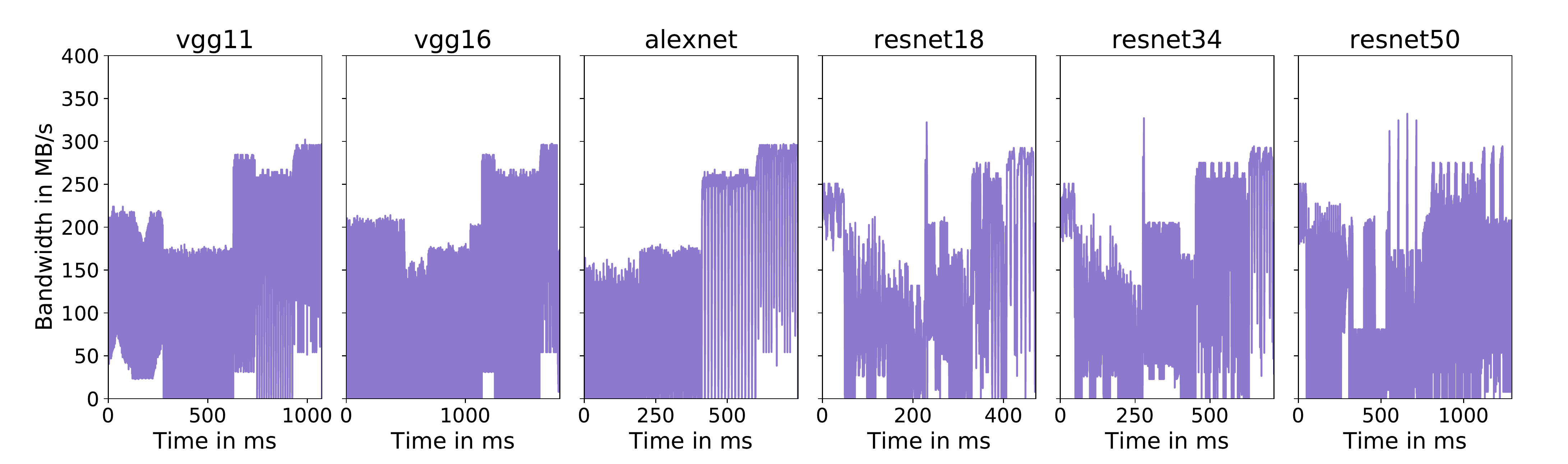}
    \caption{\textbf{The bandwidth variations on the DRAM interface during inference execution of different ML models. }}
    \label{fig:bw_variation}
\end{figure*}

We run six image classification DNN inference models,
namely, VGG-11, VGG-16, Alexnet, Resnet-18, Resnet-34 and Resnet-50.
The memory utilization shape for each of the workload is shown in Figure~\ref{fig:bw_variation}.
These traces are taken with the NPU connected to system memory bus. 
To validate possibility of attack on the LLC interface, we have also collected traces with NPU connected to LLC through accelerator coherency port (ACP). 
Those traces also have similar demand signature and are not shown due to space constraints.

For layer boundary detection, we use precision and recall
to evaluate the detector performance. Precision indicates
the fraction of correctly detected boundaries
out of the total detected boundaries.
The higher the precision is, the less false positive claims
the detector makes. 
Recall measures
the fraction of correctly detected boundaries
out of the all true boundaries.
The higher the recall is, the less correct layer boundaries
the detector misses.
To evaluate layer type classification performance,
we use accuracy which indicates the ratio of correctly
classified samples out of all samples.
We did not use accuracy for layer boundary detector because
the number of non-boundary inputs outnumbers
the number of boundary inputs by orders of magnitude.
Using accuracy may present a falsely satisfying result.

\subsubsection{\textbf{Layer Boundary Detection}}
\label{sec:boudarydetector}
A single DNN model consists of multiple layers.
The first step in the attack is layer boundary detection. 
Prior arts~\cite{ReverseCNN,DeepSniffer} demonstrated that
the read-after-write (RAW) pattern on the
\textit{address} trace reveals the layer boundaries accurately.
However, our threat model restricts
attackers to use bandwidth utilization variation rather than the address trace.


Fine-grained observations enable attackers to model each layer
as a time-series of bandwidth information.
To identify layer boundaries,
the adversary clusters the observed time-series trace
into different classes.
First, we collect
statistics like total data transferred per sampling window,
median and peak bandwidth, bandwidth standard deviation
as well as frequency domain signals extracted using
discrete wavelet transform (DWT) to perform feature extraction and selection.
Second, 
the adversary builds a bag-of-words model with the extracted features,
a popular Natural Language Processing (NLP) technique,
on sliding windows of the trace.
The combination of a bag-of-words model and sliding windows enables
the attacker classifier to obtain both the frequency- and time-domain information
of the collected trace.
Then, the attacker performs clustering to obtain
the potential layer boundary candidates.
Subsequently, offline profiled termination timings of all possible layers
are used to validate these candidates to reduce false positives.

The boundary detection results are shown in
Table~\ref{tbl:membw-channel-fpres}.
The table heading lists the different benchmarks with the total number of layer boundaries.
Adjacent layers with a different shape (\textbf{T1} in~\ref{sec:example_variation}) are identified as 
`easy'
because they utilize different memory bandwidth and are therefore easy to detect. 
{\bf The classifier can detect easy boundaries with 100\% precision for AlexNet, VGG11 and VGG16.}
\textbf{Overall, $\textbf{73.9\%}$ of layer boundaries are detected across our six workloads.}
Note that for ResNet models, the residual layers are very short,
which makes boundaries hard to detect due to our choice of 
sliding windows size.
We leave these hyperparameter tuning (e.g. sliding window size) for future work. 

\newcommand{\ra}[1]{\renewcommand{\arraystretch}{#1}}

\begin{table}[t]
\resizebox{\linewidth}{!}{%
\ra{1.4}
\begin{tabular}{ccccccccccccc}
\hline
\multirow{3}{*}{} 
&
\multicolumn{2}{c}{AlexNet}  & \multicolumn{2}{c}{VGG11} & 
\multicolumn{2}{c}{VGG16}    & \multicolumn{2}{c}{ResNet18} & 
\multicolumn{2}{c}{ResNet34} & \multicolumn{2}{c}{ResNet50} \\ 
\cline{2-13} 
&
\begin{tabular}[c]{@{}c@{}}easy\\ 3\end{tabular} & 
\begin{tabular}[c]{@{}c@{}}all \\ 4\end{tabular} & 
\begin{tabular}[c]{@{}c@{}}easy\\ 5\end{tabular} & 
\begin{tabular}[c]{@{}c@{}}all \\ 6\end{tabular} & 
\begin{tabular}[c]{@{}c@{}}easy\\ 8\end{tabular} & 
\begin{tabular}[c]{@{}c@{}}all \\ 11\end{tabular} & 
\begin{tabular}[c]{@{}c@{}}easy\\ 22\end{tabular} & 
\begin{tabular}[c]{@{}c@{}}all \\ 23\end{tabular} & 
\begin{tabular}[c]{@{}c@{}}easy\\ 24\end{tabular} & 
\begin{tabular}[c]{@{}c@{}}all \\ 36\end{tabular} & 
\begin{tabular}[c]{@{}c@{}}easy\\ 50\end{tabular} & 
\begin{tabular}[c]{@{}c@{}}all \\ 52\end{tabular} \\
\hline
\hline
\begin{tabular}[c]{@{}c@{}}precision\end{tabular} &
1 & 1 & 1 & 1 & 1 & 1 & 0.69 & 0.64 & 0.66 & 0.72 & 0.33 & 0.33 \\
\begin{tabular}[c]{@{}c@{}}recall\end{tabular} &
0.75 & 1 & 0.83 & 1 & 0.73 & 1 & 0.96 & 1 & 0.67 & 1 & 0.96 & 1\\ \hline

\end{tabular}%
}
\caption{\textbf{Precision and recall when identifying layer boundaries for each network.
{\it easy} boundaries refer to ones between
adjacent layers with different utilization shapes,
while {\it all} includes all layer boundaries.
}}
\label{tbl:membw-channel-fpres}
\end{table}

\subsubsection{\textbf{Layer Type Classification}}
\label{sec:typeclassifier}
Similar to layer boundary detection,
we include frequency domain signals from DWT to capture tile bandwidth signatures for layer type classification.
DWT detects change in bandwidth across tiles at layer boundary.
The wavelets also captures sharp changes in bandwidth,
which is useful for tile boundary detection.

We test victim memory traffic time-series using three classifiers:
Support Vector Machine (SVM),
Multilayer Perceptron (MLP), and Convolutional Neural Network (CNN),
each trained on either the time-series features for potential layers or on the features extracted from DWT.

\newcommand{\mnrow}[2][c]{%
      \begin{tabular}[#1]{@{}c@{}}#2\end{tabular}}
      
\begin{table}[tb]
\centering
\resizebox{0.95\linewidth}{!}{%
\begin{tabular}{cccccccc}\toprule
    & AlexNet & VGG11 & VGG16 & ResNet18 & ResNet34 & ResNet50 & Overall \\
\midrule
\mnrow{Execution \\time only}&
1 & 1 & 0.958  & 0.896 & 0.851 & 0.824 & 0.826 \\
\midrule
\mnrow{SVM w/ \\(w/o) DWT} &
1 & 1 & 1 & 1 & 1 & \mnrow{0.811} & \mnrow{0.927} \\ 
\midrule
\mnrow{MLP w/ \\(w/o) DWT} &
1 & 1 & 1 & 1     & \mnrow{1\\(0.986)} & \mnrow{0.868\\(0.849)} & \mnrow{0.949\\(0.934)} \\  
\midrule
\mnrow{CNN w/ \\(w/o) DWT} &
1 & 1 & 1 & 1     & 1 & \mnrow{0.877\\(0.830)} & \mnrow{0.953\\(0.934)}\\ 
\bottomrule
\end{tabular}%
}
\caption{\label{tbl:layerClassificationAccuracy}%
\textbf{Layer type classification accuracy using unshaped traffic assuming perfect layer boundary detection.
The last three rows also show results without DWT signals in parentheses.}}
\vspace{-1em}
\end{table}

Table~\ref{tbl:layerClassificationAccuracy} shows the layer-type classification accuracy for each tested model,
using bandwidth trace assuming \textit{perfect} identification of layer boundaries.
The last column (Overall) summarizes the weighted accuracy
of all layers in all models.
The first row shows classification accuracy merely using the termination timing of each layer.
\textit{This is a baseline accuracy for any attacker with the knowledge of the layer boundaries (execution timing for each layers).}
All the layers of AlexNet and VGG-11 are identifiable using this basic classifier. These workloads have few layers and all the layers differ in their execution time. Therefore, merely having the layer termination time is sufficient to classify the layers. 
The accuracy decreases for deeper models like ResNets with identical adjacent layers (\textbf{T3} in~\ref{sec:example_variation}).
The 2-4th rows show accuracy of the three evaluated layer-type classifiers.
{\bf  From execution-time based classifier to bandwidth-based classifiers,
the accuracy jumps from 84\% to 93\% on average.
From SVM to CNN, accuracy improves with increasing classifier complexity.
In addition, including frequency domain signals improves the classifier, resulting in an
accuracy of 95.3\%.
}
This is because different tile size configurations have different compute/bandwidth ratios.

\section{Countermeasures}
Bandwidth utilization channel leaks because of the 
layer dependent bandwidth utilization during ML inference.
This can be prevented
by disabling tile-size optimization or
shaping the shared interface traffic (via pure software or software-hardware co-design).

\begin{figure}[b]
    \vspace{-1.5em}\centering
    \includegraphics[width=0.8\linewidth]{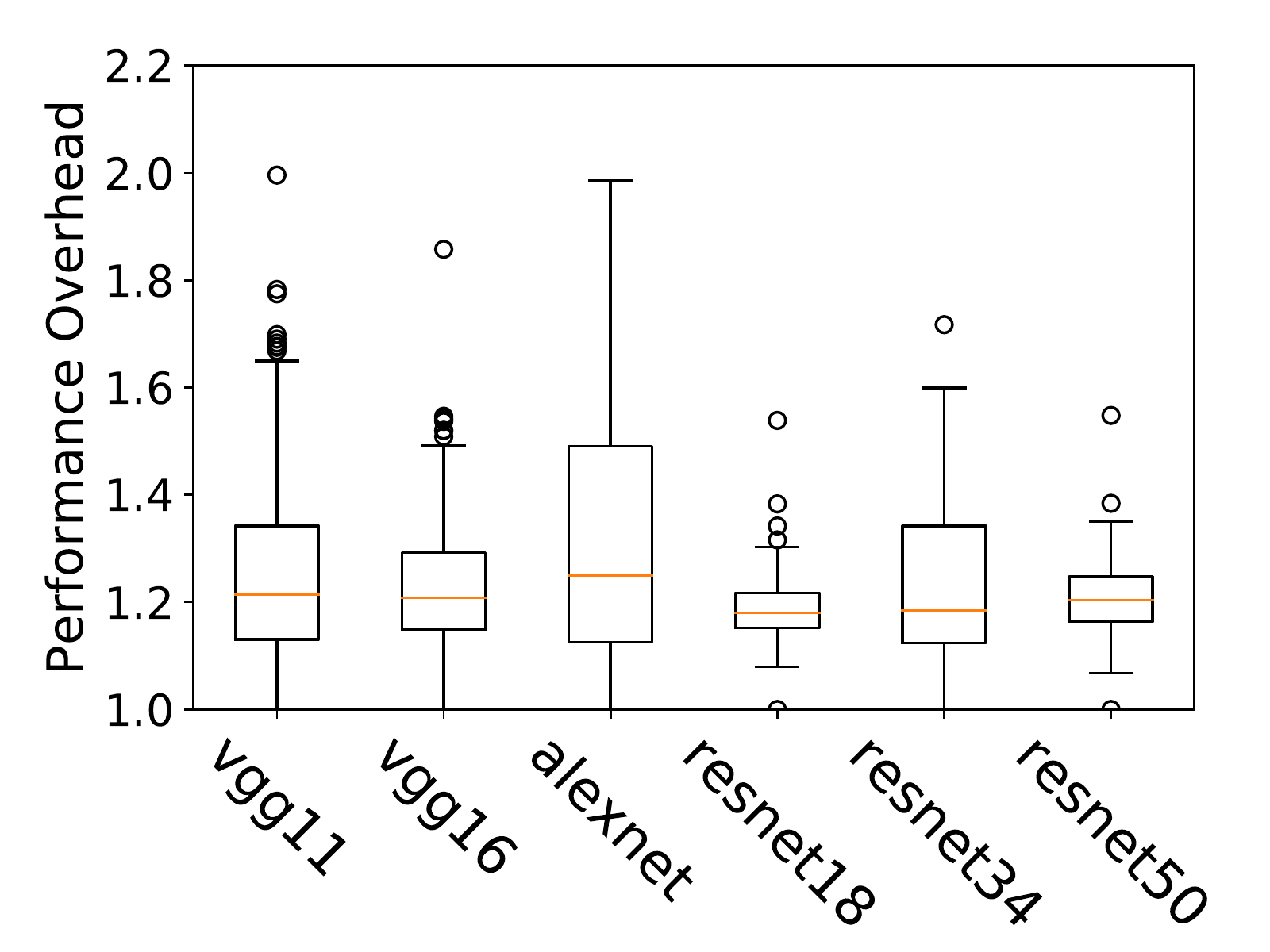}
    \caption{\textbf{Box plot showing performance variation of 800 tile size configuration for each workload }}
    \label{fig:tile_size}
\end{figure}

\subsection{Disabling tile size optimization}
Constant tile size across all layers within a model can make
the effective bandwidth constant throughout the model execution.
However, disabling this optimization leads to
inference time overhead and lower utilization of NPU resources including
compute, on-chip storage, and off-chip memory bus bandwidth.
To illustrate the wasted performance opportunity,
we explore execution performance of
800 distinct tile-size configurations for each of the models.
The overhead of each configuration with respect to the most performant
configuration is plotted  in Figure~\ref{fig:tile_size}.
\textbf{The performance overhead of the median configuration is, on an average, $\textbf{1.2}$X more compared to the best case.
The performance overhead varies up to $\textbf{1.6 - 2.0}$X in certain tile configurations.}
The resulting performance overhead
impacts the service-level agreement (SLA) and causes under-utilization of cloud resources.
This is due to the large variation in the height,
width and number of channels for different layers of the DNN model.



\subsection{Memory traffic shaping}
The memory trace could be made independent of the demand trace.
The CSP can choose a predefined memory utilization
independent from the execution model. The memory controller interleaves 
the real read requests with fake transactions if the demand falls below the expected bandwidth.
On the other hand, demand requests need to be throttled when utilization exceeds the assigned bandwidth.
The predefined memory trace can be a fixed bandwidth trace or some other model demand independent pattern.
The write requests can be sent at regular intervals. Encrypted stores with valid flag can distinguish between real and fake transactions.

To defend against bandwidth utilization attack on ML inference, we modified the accelerator design to split each transaction to multiple transactions of fixed size. The fixed size transactions were sent at an equal interval of time resulting in a constant bandwidth. Equal-sized splits are created by padding the unequal transactions.
In case of memory idle intervals, fake transactions are sent to fill the gap. The tile size optimization is performed based on the fixed bandwidth to obtain the best tile configuration for each layer.

The precision of layer boundary detection reduces drastically as shown in Table~\ref{tbl:membw-channel-const}. The boundary detection classifier has increased sensitivity to have enough coverage as visible from the recall numbers. \textit{NA} in the table for ResNet models indicate the attacker's failure to identify the true boundaries even with 10000x false positives $(precision < 0.0001)$. The layer construction is infeasible with such high false positives for the shaped constant trace. \textbf{The overall precision drops to less than $\textbf{0.01\%}$}. The model type classifier fails with such low precision on layer boundary detection thwarting the attack.

With a specific bandwidth choice, the tradeoff is between the amount of wasted memory bandwidth
and the performance overhead caused by request throttling. \textbf{With a fixed bandwidth of 200 MB/s, the geomean overhead is $\textbf{14.6\%}$ with worst case overhead of $\textbf{19.3\%}$.}

\begin{table}[tb]
\resizebox{1.01\linewidth}{!}{%
\ra{1.4}
\begin{tabular}{cccccccccccccc}
\hline
\multirow{3}{*}{} &
\multicolumn{2}{c}{AlexNet}  & \multicolumn{2}{c}{VGG11} & 
\multicolumn{2}{c}{VGG16}    & \multicolumn{2}{c}{ResNet18} & 
\multicolumn{2}{c}{ResNet34} & \multicolumn{2}{c}{ResNet50} \\ 
\cline{2-13} 
 &
\begin{tabular}[c]{@{}c@{}}easy\\ 3\end{tabular} & 
\begin{tabular}[c]{@{}c@{}}all \\ 4\end{tabular} & 
\begin{tabular}[c]{@{}c@{}}easy\\ 5\end{tabular} & 
\begin{tabular}[c]{@{}c@{}}all \\ 6\end{tabular} & 
\begin{tabular}[c]{@{}c@{}}easy\\ 8\end{tabular} & 
\begin{tabular}[c]{@{}c@{}}all \\ 11\end{tabular} & 
\begin{tabular}[c]{@{}c@{}}easy\\ 22\end{tabular} & 
\begin{tabular}[c]{@{}c@{}}all \\ 23\end{tabular} & 
\begin{tabular}[c]{@{}c@{}}easy\\ 24\end{tabular} & 
\begin{tabular}[c]{@{}c@{}}all \\ 36\end{tabular} & 
\begin{tabular}[c]{@{}c@{}}easy\\ 50\end{tabular} & 
\begin{tabular}[c]{@{}c@{}}all \\ 52\end{tabular} \\
\hline
\hline
\begin{tabular}[c]{@{}c@{}}precision\end{tabular} &
0.03 & 0.03 & 0.01 & 0.01 & 0.0027 & 0.00011 & NA & NA & NA & NA & NA & NA \\  \\
\begin{tabular}[c]{@{}c@{}}recall\end{tabular} &
0.75 & 1 & 0.83 & 1 & 0.73 & 1 & NA & NA & NA & NA & NA & NA \\ \hline
\end{tabular}%
}
\caption{\textbf{Memory traffic shaping with constant bandwidth reduces the precision of layer boundary detection drastically for all models}}
\label{tbl:membw-channel-const}
\vspace{-1em}
\end{table}

\section{Related work}

Recent works~\cite{ReverseCNN,DeepSniffer} illustrate that memory access patterns reveal a DNN model structure by snooping the off-chip address bus.
We demonstrate a new alternative side-channel,
using bandwidth variation,
for leaking model dimensions even in the presence of data and address encryption.
Observing bandwidth variation is feasible
even by on-chip unprivileged malicious co-tenants
without the use of performance counters.

Memory traffic shaping as a defence mechanism is illustrated in prior works like MITTS~\cite{mitts} or camouflage~\cite{camouflage}. These are applicable to general-purpose workloads with a runtime shaping logic.
However, the memory demand requests for a DNN workload are known at compile time.
We demonstrate that the compiler can choose a traffic pattern
and perform the tile size analysis to improve the overall inference latency. 
\section{Conclusion}

This paper studies the bandwidth utilization side-channel to infer confidential model structure.
The channel can be observed by performance counters or even by unprivileged co-executing tenant through traffic contention.
This work shows that model structure can be leaked even with an encrypted address trace.
The study discusses potential countermeasures and highlights one that
leverages the knowledge of the inference workload at compile time
to tune the tiles accordingly and improve the bus utilization
while closing the bandwidth utilization channel.

\section{Acknowledgements}
We would like to thank our anonymous reviewers for providing their valuable feedback and suggestions. This work is funded by under task $\#2965.001$ by Semiconductor Research Corporation (SRC) and Intel Strategic Research Alliance (ISRA) grant.


\bibliographystyle{IEEEtranS}
\bibliography{refs.bib}

\end{document}